\documentclass[%
 reprint,
 amsmath,amssymb,
 aps,
]{revtex4-2}

\usepackage{tabularx}
\usepackage{graphicx}
\usepackage{dcolumn}
\usepackage{bm}

\usepackage{color}
\usepackage{hyperref}
\begin{document}

\preprint{APS/123-QED}

\title{Machine Learning Enhances Algorithms for Quantifying Non-Equilibrium Dynamics in Correlation Spectroscopy Experiments to Reach Frame-Rate-Limited Time Resolution }

\author{Tatiana Konstantinova}
\author{Lutz Wiegart}
\author{Maksim Rakitin}
\author{Anthony M DeGennaro}
 \email{adegennaro@bnl.gov}
\author{Andi M Barbour}
 \email{abarbour@bnl.gov} 
\affiliation{%
 Brookhaven National Laboratory\\
 Upton, NY 11973, USA
}%

\date{\today}

\begin{abstract}
Analysis of X-ray Photon Correlation Spectroscopy (XPCS) data for non-equilibrium dynamics often requires manual binning of age regions of an intensity-intensity correlation function. This leads to a loss of temporal resolution and accumulation of systematic error for the parameters quantifying the dynamics, especially in cases with considerable noise. Moreover, the experiments with high data collection rates create the need for automated online analysis, where manual binning is not possible. Here, we integrate a denoising autoencoder model into algorithms for analysis of non-equilibrium two-time intensity-intensity correlation functions. The model can be applied to an input of an arbitrary size. Noise reduction allows to extract the parameters that characterize the sample’s dynamics with temporal resolution limited only by frame rates. Not only does it improve the quantitative usage of the data, but it also creates the potential for automating the analytical workflow. Various approaches for uncertainty quantification and extension of the model for anomalies detection are discussed.
\end{abstract}

\maketitle


\section{\label{sec:Introduction}Introduction}

\begin{figure*}[]
\includegraphics[width=\linewidth]{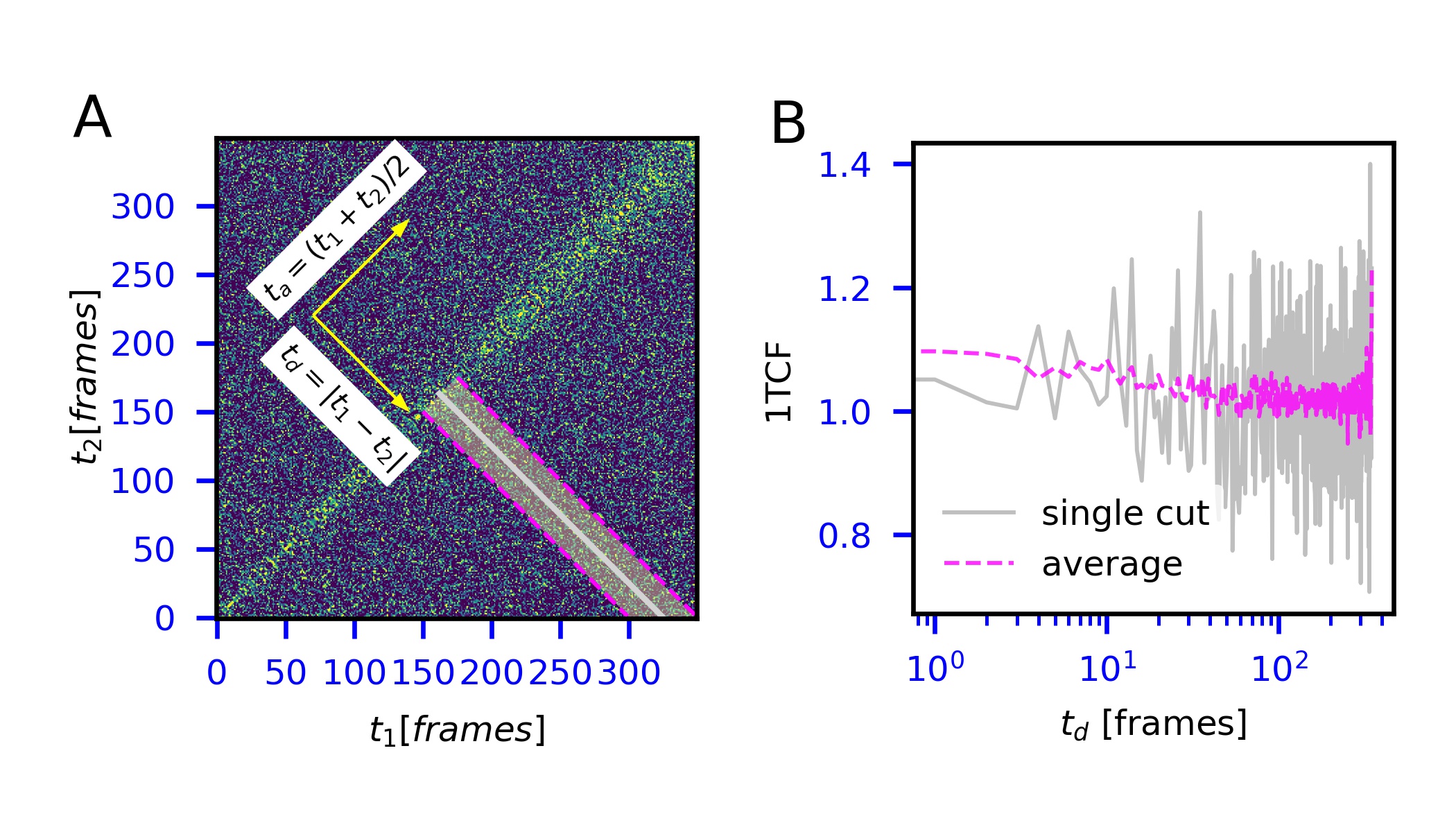}%
\caption{\label{fig:Figure1}Examples of intensity-intensity correlation functions. (A) A non-equilibrium $2TCF$. $t_1$ and $t_2$ are experiment time measured in frames. Yellow arrows indicate the directions of age axis \emph{$t_a$} and time delay axis \emph{$t_d$}, defined according to the formulas in the inset.  Pink dashed lines and the grey solid line represent the edges and the bin center, respectively, for the region used for generating the $1TCF$ in (B). (B) $1TCF$s for a single cut along the bin center (grey solid) and average within the bin of 50 frames(pink dashed).}
\end{figure*}

Technological advances, such as high-brightness X-ray photon sources \cite{bergmann_2009, couprie_2014, vartanyants_2016} and high-rate high-sensitivity detectors \cite{Grybos_2016, Llopart_2002}, enable new discoveries by means of X-ray Photon Correlation Spectroscopy (XPCS) \cite{zhang_2018}. Non-equilibrium dynamics are signature of many systems studied with XPCS, including superconductors \cite{ricci_2020}, metallic glasses \cite{evenson_2015}, proteins \cite{begam_2021} and polymeric nanocomposites \cite{yavitt_2020}. Such dynamics are commonly represented via two-time intensity-intensity correlation functions ($2TCF$s) \cite{brown_1997}, defined by the expression: 
\begin{equation}\label{eq:(1)}
C2(\pmb{q};t_{1}, t_{2}) = \frac{\langle I(\pmb{q};t_{1})I(\pmb{q};t_{2})\rangle}{\langle I(\pmb{q};t_{1})\rangle \langle I(\pmb{q};t_{2})\rangle}
\end{equation} 
where \(I(\pmb{q};t)\) is the intensity of a detector pixel corresponding to the wave vector \(\pmb{q}\) at time \(t\). The average is taken over pixels with equivalent \(\pmb{q}\) values.
The quantitative analysis of a non-equilibrium $2TCF$ (Fig.~\ref{fig:Figure1}(A)) often starts with its binning along the sample’s age axis \emph{$t_a$} into quasi-equilibrium regions \cite{madsen_2020, bikondoa_2017} prior to making cuts along the time delay axis \emph{$t_d$} and averaging them, i.e. obtaining ‘aged’ one-time intensity-intensity correlation functions ($1TCF$s)\cite{madsen_2010} of delay time:
\begin{equation}\label{eq:(2)}
C1(\pmb{q},t_a;t_d) = {\langle C2(\pmb{q};t_a, t_d)\rangle}_{t_a}
\end{equation}
The reason for binning and averaging is increasing the accuracy of further analysis. The resulting $1TCF$s are fit to a functional form that is characteristic for the dynamics under investigation.

The traditional analysis of experimental data with non-equilibrium dynamics presents several challenges. Selection of the quasi-equilibrium regions is often a non-trivial and time-consuming task, which requires interventions from a human researcher for a visual inspection of the calculated $2TCF$ and intermediate results. Moreover, the temporal resolution of the evolution of dynamics’ parameters deteriorates due the binning procedure. Besides, with increasing data collection rates, it often becomes unfeasible to properly do on-the-fly data analysis during XPCS experiments when such analysis requires inputs from a researcher. This can lead to inefficient use of beam facilities and the efforts of scientific stuff as well as to unfortunate omission of important parameter space regions during measurements. The sheer volume of collected data can reach hundreds of thousands of $2TCF$s a day, making even the offline data processing a challenging task. Automating the analytical routine is an inevitable step towards high-throughput autonomous XPCS experiments. Ability to perform the analysis with experimental frame-rate resolution is the ultimate target.

Since the motivation behind binning and averaging the data is to curtail the impact of noise, pre-processing a $2TCF$ with a noise-reducing algorithm would increase the chance of achieving a frame-rate resolution along \emph{$t_a$} during quantitative analysis. Previously, we demonstrated an approach for noise reduction in $2TCF$s with equilibrium dynamics, which is based on Convolutional Neutral Network--Encoder-Decoder model (CNN--ED) \cite{konstantinova_2021}. The method is shown to work well for equilibrium data and has a promising potential for use with slowly ageing dynamics when applied as a sliding window along \emph{$t_a$}. However, there are two major concerns regarding the use of the model for general cases of non-equilibrium dynamics. Firstly, if the rate of the dynamics changes fast, an equilibrium approximation may not be valid, even in narrow age regions. Thus, the application of the model would lead to loss of experimental resolution and extraction of less accurate results. Secondly, for a $2TCF$ containing a much larger number of frames than the input size of the model, a considerable portion of it would still contain the original level of noise after the model is applied along \emph{$t_a$}. 

Here, we build upon the previous model to address the reduction of noise in arbitrary-sized $2TCF$s with non-equilibrium dynamics such as ageing. We further demonstrate how the new model, integrated into a workflow for quantitative analysis of XPCS data, eliminates the need of age binning and allows wider parameter bounds for the fit of $1TCF$s. The results of such analysis have temporal resolution, which is limited only by the experimental acquisition rates. The methods for estimating the creditability of the results -- uncertainty scoring for denoised $2TCF$s and \emph{trust regions} for dynamics’ parameters -- are discussed. Various options for analysis workflows are demonstrated for several XPCS experiments. 

\section{\label{sec:Methods}Methods}

\subsection{\label{sec:Data}Data}

For a denoising model to perform well for non-equilibrium $2TCF$s, data from experiments with these types of dynamics are used for training. The data are collected from 65 XPCS experiments at the CHX beamline of NSLS-II from three different samples, recorded at different conditions and with different detector collection rates. Some of the measurements are repeated multiple times at the same conditions. The samples exhibit dynamics, common for many materials, that at various rates monotonically accelerate with age, monotonically decelerate with age or stay at quasi-equilibrium. 
For all considered dynamics, individual $1TCF$s can be approximated with the Kohlrausch-Williams-Watts (KWW) form:
\begin{equation}\label{eq:(3)}
C1(t_d) = C_{\infty} + \beta e^{-2(\Gamma t_d)^{\alpha}}
\end{equation}
where $\Gamma$ is the rate of the dynamics, $\beta$ is the contrast factor, $\alpha$ is the compression constant and $C_{\infty}$ is the baseline. All parameters are functions of $t_a$ and $\pmb{q}$.
Several regions of interest in the reciprocal space are considered for each experiment when calculating $2TCF$s. There is a total of 492 $2TCF$s, ranging from 134 to 2950 frames. The original full-sized $2TCF$s were split between training (454) and validation (38) sets prior to generating the inputs to the model and augmenting the data as described below.

Inputs are obtained from raw experimental $2TCF$s by cropping the 100$\times$100 frames pieces along the $t_a$ in a similar way as was done previously \cite{konstantinova_2021}. Prior to cropping, the diagonal values, containing high errors, are replaced by the average of their nearest off-diagonal neighbors. That is, the value at [j, j] is replaced by the mean of the values at [j-1, j] and [j+1, j]. This approach preserves the information in the experiments with slowly varying contrast. For data augmentation, additional inputs are constructed by considering every 2\textsuperscript{nd}, 3\textsuperscript{rd}, 4\textsuperscript{th} and 5\textsuperscript{th}  frame of the original data when enough frames are available to form at least one 100$\times$100 example. The final size of the training and validation sets are 25222 and 1557 cropped examples respectively.  
Each input is scaled to have zero mean and unit variance prior to passage through the model. A reverse transformation is performed for the model output. 

\subsection{\label{sec:Denoising Autoecoder Model}Denoising Autoecoder Model}

The CNN-ED model architecture, used for equilibrium data, demonstrated several advantages, such as simplicity, control of overfitting as well as fast training and application. Trying the same model architecture with certain adjustments for non-equilibrium data is a natural choice. The model presented here consists of an encoder with two 10-channel convolutional layers, 8-dimensional latent vector and the decoder with two 10-channel transpose-convolutional layers. Two modifications are implemented during the model training to meet the peculiarities of non-equilibrium dynamics. Firstly, since a ‘noise-free’ version of an input cannot be obtained by averaging multiple cropped inputs collected from the same full-sized $2TCF$, the model is trained in an autoencoder (AE) mode with an input and its target being the same. Secondly, a non-equilibrium $2TCF$ cannot be described by a single $1TCF$ and thus only the mean square loss between the output and the target are chosen as the training loss function.  

Despite the increased complexity of dynamics in the training data, the architecture still appears appropriate for the denoising task based on the model’s good performance for the validation set [see Appendix~\ref{sec:Different Kernel Sizes}, \ref{sec:Model Variance}].  The effectiveness of models with convolutional kernel sizes from 1 to 17 is tested and only minor improvements in the smoothness of the output and the level of finite details preserved for larger kernels are discovered. The models with larger kernels take longer to train and the computational time during their application also becomes significant. Further in the text, unless stated otherwise, a model with the kernel size 1 is considered. The fact that the same model architecture works well for both equilibrium and non-equilibrium ageing $2TCF$s confirms it robustness for different types of dynamics.

\subsection{\label{sec:Adapting Model to Arbitrary Input Size}Adapting Model to Arbitrary Input Size}

Since the model has a fully connected layer (the latent vector), it can only be applied to a fixed-size input, i.e. 100$\times$100 frames. However, the dimensions of an experimental $2TCF$ can vary considerably. The symmetry of a $2TCF$ prevents the model to be applied as a raster scan over the entire image. Application of the model in a sliding window fashion along the age axis $t_a$ can help with the noise reduction in certain cases like in out original quasi-equilibrium analysis~\cite{konstantinova_2021}.
Nevertheless, this approach is not suitable for general cases of $2TCF$s, where dynamics may happen outside of the first 100 delay frames along the $t_d$ axis.
 
\begin{figure*}
\includegraphics[width=\linewidth]{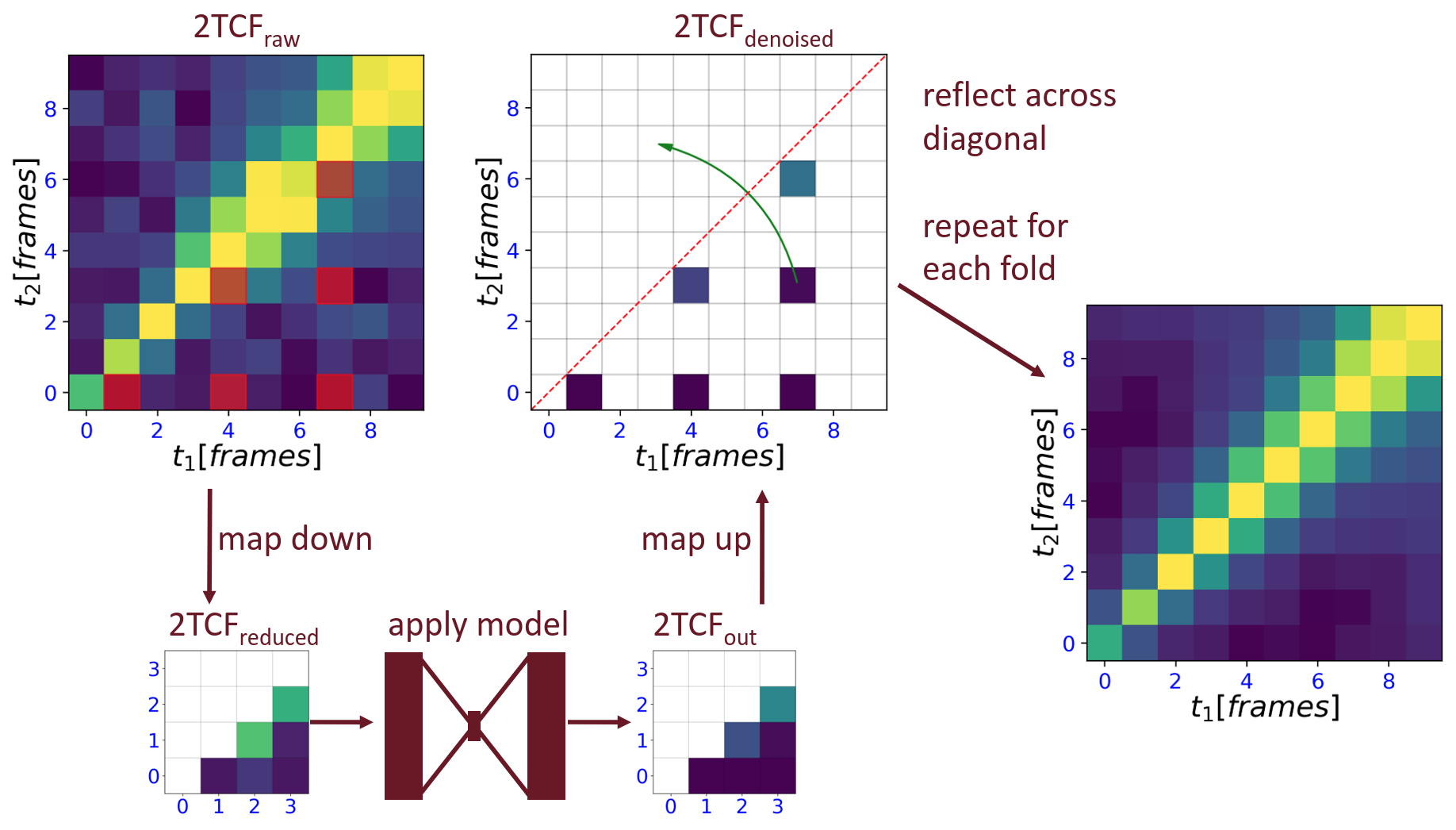}%
\caption{\label{fig:Figure2} Scheme of the model application to an input of arbitrary size.}
\end{figure*}

Figure~\ref{fig:Figure2} shows a possible solution, which is inspired by the procedure of data augmentation when the original $2TCF_{raw}$ is mapped down (\emph{down-mapped}) to a smaller size by considering every N-th frame horizontally and vertically, where N = 2, 3, 4, 5. For model application, N is a whole part of division of total number of frames in $2TCF_{raw}$ by 100. This way, the size of a newly constructed $2TCF_{reduced}$ matches the required model’s input size. By its definition (Eq.~\ref{eq:(1)}), a $2TCF$ is symmetric with respect to $t_a$. However, when the starting point for \emph{down-mapping} is away from the $t_a$ axis ($t_d>$0) of $2TCF_{raw}$, the reduced version is not symmetrical. To ensure the symmetry of the input, only the bottom part of the $2TCF_{reduced}$ is considered and a reflection across the $t_d$=0 line is applied to it. For all starting points, the  $2TCF_{reduced}$ values at $t_d$=0 are extrapolated by the neighboring points in the same way it is done for model’s training. After application of the model, its output $2TCF_{out}$ is mapped back (\emph{up-mapped}) to the size of the $2TCF_{raw}$ by placing [i, j]-th point from the $2TCF_{out}$ to [N*i, N*j]-th point in the resulting denoised $2TCF_{denoised}$. Again, only the bottom half is filled and then reflected across the $t_d$=0 line. The process is repeated for all possible starting points, covering fully the $2TCF_{raw}$.

For regions of very fast dynamics and for large-sized $2TCF$s, this \emph{down-and-up mapping} approach may impair the experimental temporal resolution. In such situations, the values near the diagonal of $2TCF_{raw}$ can instead be processed in a sliding-window fashion, when the model is successively applied to parts consisting of cropped 100$\times$100 frames equally distanced along ta, and the outputs are averaged. The results of the model’s application in the \emph{down-and-up mapping} and sliding window fashion are combined to obtain the final denoised version of the original $2TCF_{raw}$. This method reduces noise in the entire $2TCF_{raw}$ and preserves the experimental temporal resolution for both $t_a$ and $t_d$.  

\subsection{\label{sec:UQ}Uncertainty Quantification for Denoising }

When applying a model to unseen data, there are three potential sources of uncertainty: error (noise) of the input, model’s bias and model’s variance. The bias of the model primarily originates from selection of the training dataset and constricting the model’s architecture. Variance appears due to random initializations of model’s weights and ordering of data butches during model’s training. Approaches for quantifying uncertainty for a deep learning model often involve random perturbations of the model during training and/or application and obtaining the distribution of respective outputs \cite{abdar_2021}.

We estimate the variance of the denoising AE by considering deviations of outputs for models trained with different random initializations (see Appendix \ref{sec:Model Variance}). It appears that the model’s variance is generally much smaller than the corresponding values of $\beta$. Moreover, the fluctuations of values in the neighboring points of a $2TCF_{denoised}$, caused by \emph{down-and-up mapping} during model application, scale linearly with the inherent variance of the model. Thus, it is not necessary to separately estimate the model’s variance for experiments with more than 100 frames, as the point-to-point variations in a $2TCF_{denoised}$ already influence the quantitative analysis of its $1TCF$s.

From the practical point of view, the main potential source of uncertainty of the model is its bias. Obviously, a denoising AE must exhibit \emph{some} bias to remove noise. Hence, to drive the decision about applicability of the model in each case, the quantification of the bias ought to answer the question: how certain can one be that the model’s output is a valid representation of the underlying sample’s dynamics in the input? Naturally, a model returns the most trustworthy results for the inputs that are very similar to the examples in the training set. We suggest two quantitative measures -- uncertainty scores -- for estimating the bias by comparing a new input to the training examples.

The first measure is based on the Euclidean distance between the latent vector representation of an input and the center of the latent representations of all training examples. The analysis of the pairwise distributions of the 8 latent coordinates of the training set reveals that they form a single compact cluster [see Appendix \ref{sec:Latent Coordinates Distribution}]. The shape of the cluster is determined by the distribution of dynamics and the noise level in the training set. However, the values of each of the eight latent coordinates are mostly concentrated around a single point. The distance from this central point is thus indicative of how different a given input is from most of the training data and thus can be used as an uncertainty score. To ensure an equal contribution of each latent dimension to the uncertainty score, all coordinates are scaled to have a zero mean and unit variance across the training dataset. To improve the interpretability of the bias, the distance to the center is normalized by its median value among the training set examples. The normalization of the distance helps to set a general threshold for an acceptable level of the bias without reference to a particular model or the training set.

The second measure of bias of the denoising is less abstract and does not depend on the model architecture. It involves evaluation of trends in residuals, i.e., the differences between the model’s targets and corresponding outputs. When the output is a valid representation of the sample’s dynamics, the residual should resemble random and sometimes correlated noise in form of vertical and horizontal stripes, without trends along neither $t_a$ nor $t_d$. Since the training examples mostly represent monotonically ageing dynamics, the model tends to perform well for similar cases. A trend in a residual can indicate a heterogeneity on top of the ageing dynamics or a completely different type of dynamics. Such situations require additional attention during analysis. 
To identify the trends, it is convenient to look at the projection of a two-dimensional residual to the vertical (or horizontal) direction. Without trends, the autocorrelation coefficients (ACCs) of this projection are close to zero. In contrast, when a trend is present, the absolute value of the ACCs grows. We calculate the first ACC for all the examples of the training set and calculate probability density distribution using Gaussian kernel density estimator [see Appendix \ref{sec:Quantifying Bias}]. The probability density for the first ACC of the residual for a new test example is the second type of uncertainty scoring of the model prediction.

Since both bias measures are based on the distribution of the training examples, they are suitable for detecting anomalies – inputs that are different from the training set. The analysis workflow can flag a new $2TCF$ as an anomaly if its uncertainty score is above the user-defined threshold. 

\subsection{\label{sec:Extracting Dynamics Parameters}Extracting Dynamics Parameters} 

Quantitative analysis of XPCS experimental results with non-equilibrium dynamics, such as ageing, involves taking time-slices along the delay axis $t_d$ of a $2TCF$ at different sample’s ages $t_a$ and fitting the resulting $1TCF$s with an appropriate model that describes the dynamics, which in many cases is the KWW form Eq.~\ref{eq:(3)}. Typically, several adjacent slices within age range $\Delta t_a$ need to be averaged to target a signal-to-noise ratio that is sufficient for extracting dynamics parameters with reasonable certainty. It is necessary that the dynamics do not change much within $\Delta t_a$. An example of selecting a quasi-equilibrium bin within a $2TCF$ is shown in Fig.~\ref{fig:Figure1}. This approach leads to a loss of temporal resolution for the parameters. Moreover, the procedure of binning $1TCF$s often requires an repetitive evaluation of the fitting results and adjustment of bins’ boundaries by an expert researcher. For a large volume of collected data, it becomes challenging to properly split $2TCF$s for each experimental region of interest (ROI), leading to growing uncertainty of the results. Ability to perform an adequate quantitative analysis for a single value of $\Delta t_a$ across the entire experiment would facilitate the automation of the analysis process.  The highest temporal resolution and maximum usage of experimental data are achieved when considering cuts with $\Delta t_a$ = 1 $\frac{1}{frame\ rate}$ (or, $fr^{-1}$).

\begin{figure*}
\includegraphics[width=300pt]{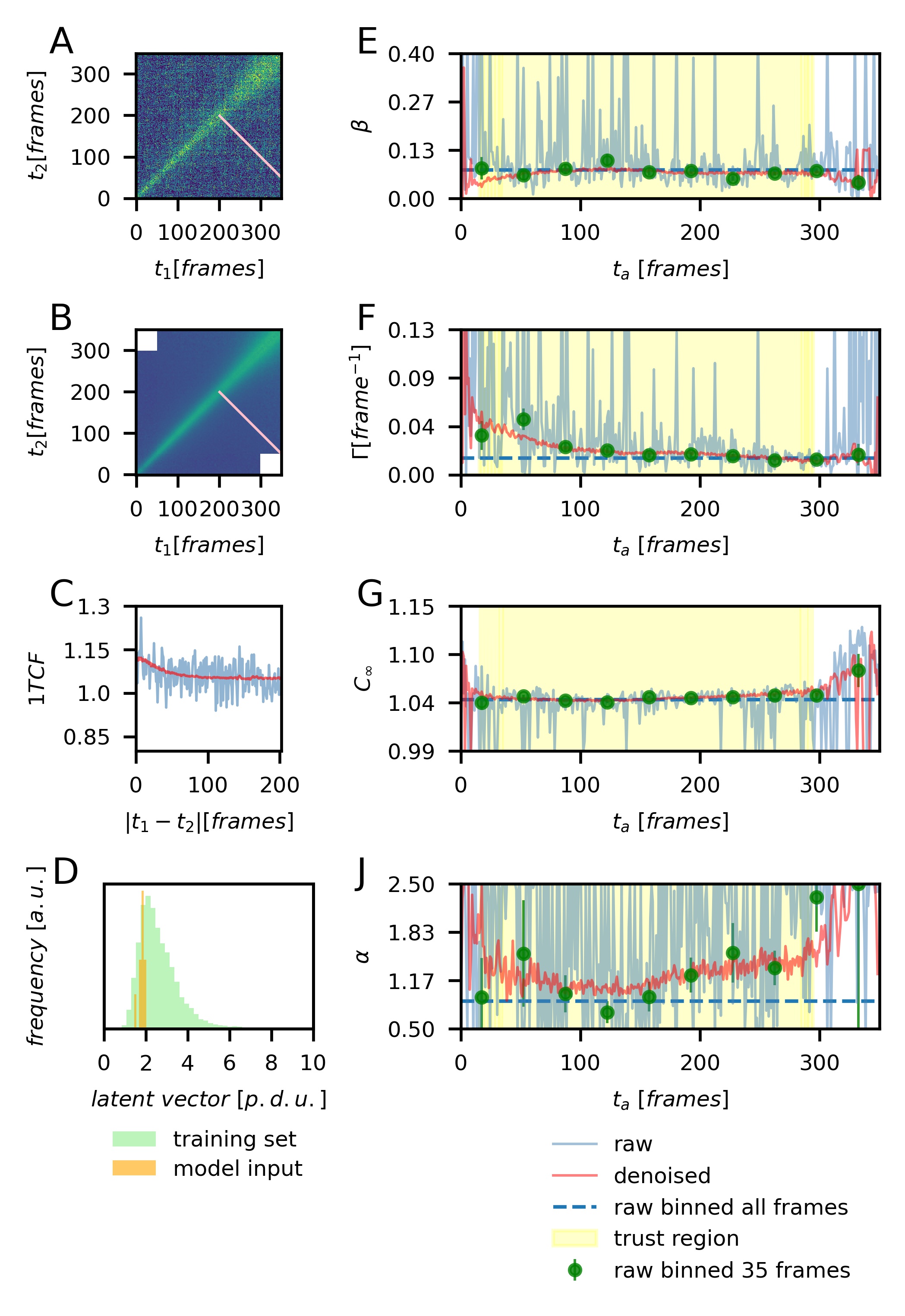}%
\caption{\label{fig:Figure3} Comparison of quantitative analysis for $2TCF_{raw}$ and $2TCF_{denoised}$. (A)$2TCF_{raw}$, (B) $2TCF_{denoised}$, (C) $1TCF$s taken along the line in (A-B), (D) distribution of latent vector lengths for the training set (green) and the $2TCF_{reduced}$s (orange), (E-J) plots for dynamics parameters values according to the legend. The \emph{trust regions} are calculated for fits of $2TCF_{denoised}$. }
\end{figure*}

We compare the possibility of conducting a quantitative analysis of a $2TCF_{raw}$ and a corresponding $2TCF_{denoised}$ while considering $1TCF$ cuts with $\Delta t_a$ = 1 $fr^{-1}$ without a prior knowledge of the dynamic's parameter ranges (Fig.~\ref{fig:Figure3}). An example of a noisy $2TCF_{raw}$ (350 frames) from the validation set (Fig.~\ref{fig:Figure3}A) is passed through the model, resulting in the $2TCF_{denoised}$ with a significantly reduced level of noise (Fig.~\ref{fig:Figure3}B-C). The latent space representations of the 100$\times$100 frames $2TCF_reduced$ are close to the center of the training set (Fig.~\ref{fig:Figure3}D), indicating that the model output is likely a valid representation of the dynamics in the input. To extract the parameters, each of the $1TCF$ cuts is fitted to Eq.~\ref{eq:(3)}. Cuts with less than 5 points are not considered. The results of the fit are shown in Fig.~\ref{fig:Figure3}(E-H). The fits for the $1TCF$ cuts from the $2TCF_{raw}$ with  $\Delta t_a$ = 35 $fr^{-1}$ are provided as ground truth values. Apparently, high noise in the $2TCF_{raw}$ does not allow extracting meaningful information about temporal evolution of dynamics parameters without restricting the fit parameters within narrow regions or increasing the bins’ width. The $2TCF_{denoised}$, on the other hand, produces clear, slowly evolving trends, matching the ground truth values.

While binning and averaging $1TCF$s reduces the noise, it is not a universal solution because material’s dynamics can vary considerably within an age bin, and therefore, a single set of parameters cannot describe them.  For the example shown in Fig.~\ref{fig:Figure3}, this becomes apparent when considering averaging all available $1TCF$s and fitting the result to Eq.~\ref{eq:(3)}. Since $\beta$ and $C_{\infty}$ are not changing in this experiment, the resulting values for these parameters are close to the ground truth values. However, the values for $\Gamma$ and $\alpha$ are not close to their corresponding average values. 

For an automatic analysis, it is important to flag the results that cannot be fully trusted, even when the goodness-of-fit to Eq.~\ref{eq:(3)} is acceptable. If dynamics are not fully captured by the experiment, parameters of the Eq.~\ref{eq:(3)} become mutually dependent and the outputs of non-linear regression can be misleading because several different sets of the dynamics’ parameters can result in almost equally good fits. Therefore, we introduce the concept of \emph{trust regions} for the parameters. A \emph{trust region} of a parameter is a binary vector with the length equal to the number of bins along $t_a$, which indicates whether the parameter is likely to be reliably identified within each bin. The binary values are determined by several criteria including the rate of the dynamics, correlations between parameters and parameters' relative errors.  

When a material’s dynamics are slow with respect to the duration of the experiment, the baseline $C_{\infty}$ cannot be reliably identified. Likewise, the true value of  $\beta$ cannot be extracted if the dynamics are not fully captured by the experiment due to them being either too fast or too slow. The quantitative measure for establishing respective thresholds for dynamics’ rate is a half-time, i.e., the time it takes for the contrast to drop by half: 

\begin{equation}\label{eq:(4)}
T_{1/2} = \left(\frac{ln2}{2} \right)^{\frac{1}{\alpha}}\frac{1}{\Gamma}
\end{equation}

For slow dynamics, when the half-time is larger than a user-defined portion of $1TCF$’s length, the \emph{trust region} values for $C_{\infty}$ and $\beta$ are set to zero: not trustable. In case of fast dynamics, when the half-time is less than a certain number of frames, the \emph{trust region} for $\beta$ is set to zero. However, the $C_{\infty}$ can be reliably extracted for fast dynamics and hence its \emph{trust region} values are set to one. Similarly, threshold-based conditions are defined for the correlation coefficient between parameters, relative errors of the parameters and $R^2$ measure of the fit.

\section{\label{sec:Results}Results}

\subsection{\label{sec:Analysis Workflow}Analysis Workflow}

\begin{figure*}
\includegraphics[width=\linewidth]{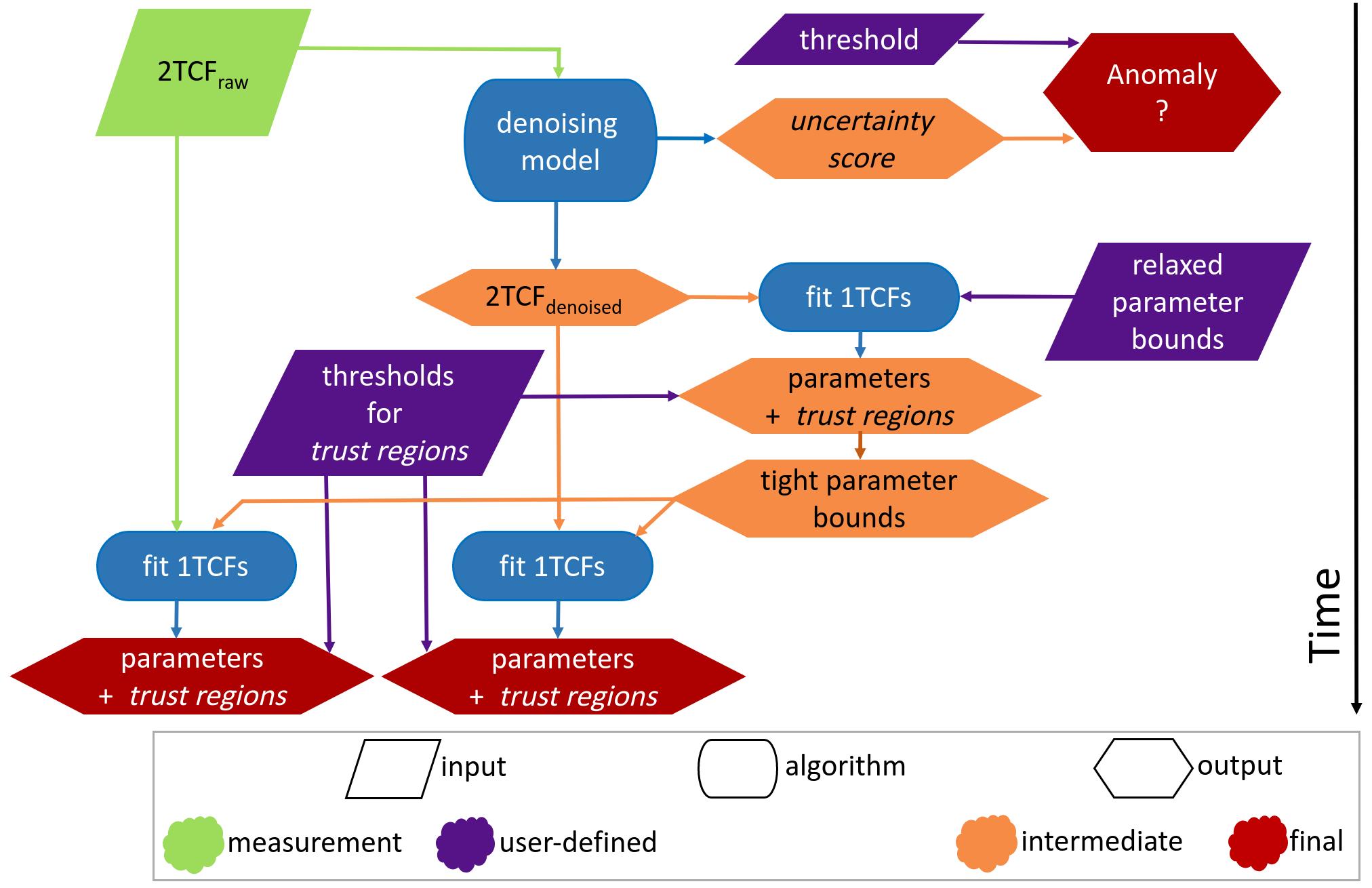}%
\caption{\label{fig:Figure4} Analysis workflow.  Shape keys: rhombus – data supplied to the algorithm, rounded shape - model, diamond shape – result of calculations. Color keys: green – results of XPCS measurements, purple – supplied by a user, orange – intermediate results, red – the final results.}
\end{figure*}

Depending on the application, there are various ways to design an analysis workflow that includes $2TCF_{raw}$s and $2TCF_{denoised}$s as well as any prior knowledge about the material’s dynamics. An example flowchart for extracting dynamical parameters from a $2TCF_{raw}$ is shown in Fig.~\ref{fig:Figure4}. In this workflow, a $2TCF_{denoised}$ is used for flagging unusual observations and narrowing parameter boundaries based on the initial fit. The new boundaries are then used to fit the $2TCF_{raw}$ and the $2TCF_{denoised}$. The last step of the analysis results in two sets of parameters with corresponding errors and \emph{trust regions}: one for the $2TCF_{raw}$ and one for the $2TCF_{denoised}$. Step-by-step dynamics extraction algorithms are provided in Appendix~\ref{sec:Algorithms}. 

Naturally, when reporting experimental outcomes, the fitting results of unprocessed $2TCF_{raw}$ have a higher priority with respect to the fitting results of $2TCF_{denoised}$ because signal processing, such as noise removal, introduces an uncertainty.  According to the presented analysis workflow, $2TCF_{denoised}$ can be used for evaluating the parameters’ boundaries for fitting the raw data during the second pass with the same or greater bin width. However, it is also possible to supplement the fitting results for $2TCF_{raw}$ with fitting results for $2TCF_{denoised}$ for regions in parameter space where the low signal-to-noise ratio in raw data prevents from obtaining meaningful parameter values. Note that removing the noise is just an additional step in extracting parameters from the raw experimental data. Therefore, the fitting results for $2TCF_{denoised}$ can be tested against $2TCF_{raw}$ with goodness-of-fit measures such as $R^2$. Alternatively, one can use the fitting results for $2TCF_{denoised}$ to select quasi-equilibrium regions for the analysis of $2TCF_{raw}$. 

Upon calculating the uncertainty score for the $2TCF_{denoised}$, one may wish to extend the workflow to investigate possible dynamic heterogeneities of the sample by relying on the denoised output and the results of the $1TCF$s' fit. The bias of the denoising model provides the opportunity to separate the average ‘envelope’ dynamics and stochastic heterogeneities by subtracting either the $2TCF_{denoised}$ or the $2TCF$, computed based on the fit parameters for the $2TCF_{denoised}$, from the original $2TCF_{raw}$.

\subsection{\label{sec:Application to New Data: Standard Analysis}Application to New Data: Standard Analysis}

We test a workflow for XPCS analysis with the use of denoising AE on previously reported data for 3D printing with solution of Lithium Titanate particles \cite{lin_2021}. The experiment is not a part of the training or the validation datasets used in the current work. However, the far-from-equilibrium dynamics of the ink, exhibited during its deposition and recovery, are similar to the types of dynamics used during the models’ training. Thus, the experiment presents an intended use case for the denoising AE. 

\begin{figure*}
\includegraphics[width=350pt]{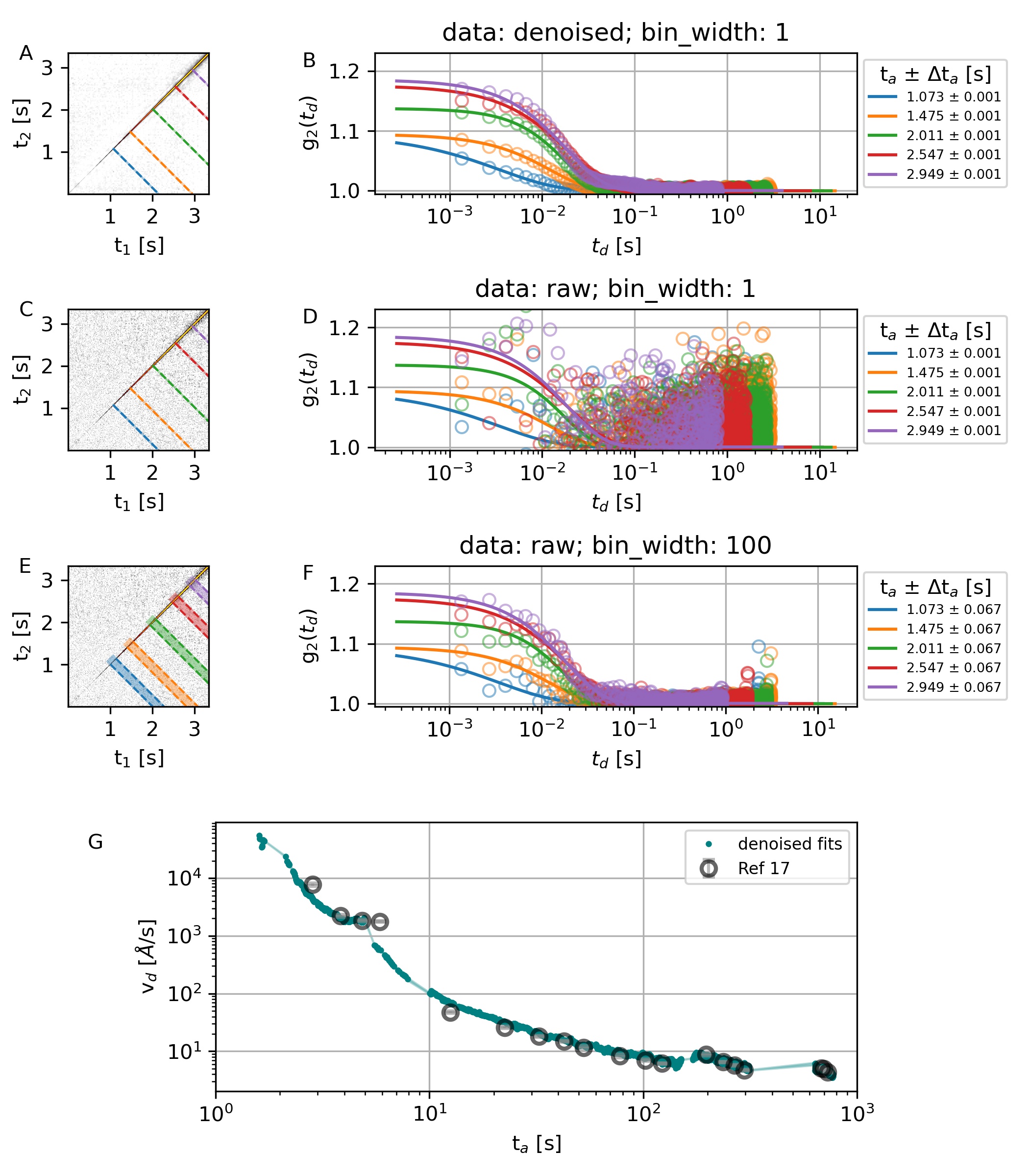}%
\caption{\label{fig:Figure5} Comparison of use $2TCF_{raw}$ and $2TCF_{denoised}$ from 3D printing experiments \cite{lin_2021} for quantitative analysis. 
(A) $2TCF_{denoised}$ and bins of width 1 frame (dashed lines).
(B) $1TCF$s corresponding to the bins in (A) (open circles) and their fits to Eq.~\ref{eq:(3)} (solid lines).
(C) $2TCF_{raw}$ and bins of width 1 frame (dashed lines). (D) $1TCF$s corresponding to the bins in (C) (open circles) and fits to denoised data from (B) (solid lines).
(E) $2TCF_{raw}$ and bins of width 100 frames (dashed lines). (F) $1TCF$s corresponding to the bins in (E) (open circles) and fits to denoised data from (B) (solid lines).
(G)drift velocity obtained from a set of experimental scattering series, calculated using $2TCF_{denoised}$ (solid teal circles) and from Ref.~\cite{lin_2021}. Results from $2TCF_{denoised}$ have 7854 points and the previous results have 19 points. }
\end{figure*}

There are several factors that limit the signal-to-noise ratio in this experiment: operando character of the measurements, beam-sensitivity of the ink and anisotropy of the dynamics, which requires selection of small ROIs in the reciprocal space for analysis. In the original work, unevenly spaced age bins of various width have been selected to obtain good quality ‘aged’ $1TCF$ slices \cite{lin_2021}. Here, we test the advantages of applying the denoising AE for conducting the same analysis with bin width $\Delta t_a$ = 1 $fr^{-1}$. 

Figure~\ref{fig:Figure5} provides an example of the denoising AE being applied to the $2TCF_{raw}$ for one of the ROIs. Since the dynamics is relatively fast, the points near the $t_a$ axis are denoised when the model is applied in a sliding window fashion and the points away from $t_a$ are denoised via \emph{down-and-up mapping} approach. The difference in level of noise for $2TCF_{raw}$ and $2TCF_{denoised}$ can be seen from corresponding $1TCF$s taken at different ages. Bin width of 1 frame has been considered for $2TCF_{denoised}$ and resulting $1TCF_{denoised}(t_a)$ are fitted to Eq.~\ref{eq:(3)} to obtain $1TCF_{fit}(t_a)$. $1TCF_{raw}(t_a)$ for bin widths 1 and 100 are compared against the same $1TCF_{fit}(t_a)$. As expected, the fit to denoised version of correlation function describes the raw data well. As the width of the bins increases, the values of $1TCF_{raw}(t_a)$ lie closer to the corresponding $1TCF_{fit}(t_a)$ lines. The model captures well not only the characteristic time of the signal decorrelation, but also the change in the contrast factor $\beta$, caused by initial fast non-ergodic dynamics with timescales outside of the experimental time window. 

The sample’s dynamics for multiple ROIs is quantified according to the scheme in Fig.~\ref{fig:Figure4}. We calculate the drift velocity from relaxation times at different wave vectors along deposition direction $\Phi = -90^{\circ}$ using only denoised $2TCF$s and compare them to the values obtained in Ref.~\cite{lin_2021}, as shown in Fig.~\ref{fig:Figure5}G. The values obtained from $2TCF_{denoised}$ are very close to results of the original analysis, confirming the denoising AE does not distort the data. Instead, it eliminates the need to carefully select the age bins and ensures the ultimate temporal resolution of a single acquisition period. 

\subsection{\label{sec:Outlier Detection}Application to New Data: Outlier Detection}

By using latent space representation of a denoising AE, it is possible to quantitatively compare $2TCF$s from the same group of measurements even if the model was not exposed to this type of dynamics during training. This is suitable for anomaly detection in series of consecutive XPCS measurements for the same material. We test the application of the model for instabilities detection during data collection for static dynamics of a LSAT sample (MTI Corporation) at the CSX beamline of NSLS-II. Six data series, 7200 frames (exposure 1.0875 seconds) each are purposefully collected during synchrotron accelerator testing to introduce a disturbance of experimental conditions. Three of the series are collected in a stable beam regime, one series is recorded with small disturbance of the beam energy and two series are recorded during strong beam energy disturbance. From each series, 72 model's inputs (size 100$\times$100 frames) are generated by considering every 72\textsuperscript{nd} frame with different starting points. Each of the \emph{down-mapped} inputs from a single series contain the same information, but different noise. To make the comparison sensitive to the mean value and variance of the contrast, the inputs are not scaled, but are clipped within values of 1 and 2. The values at $t_d$ = 0 are extrapolated with the neighboring values as it is done for a routine model application.

\begin{figure*}
\includegraphics[]{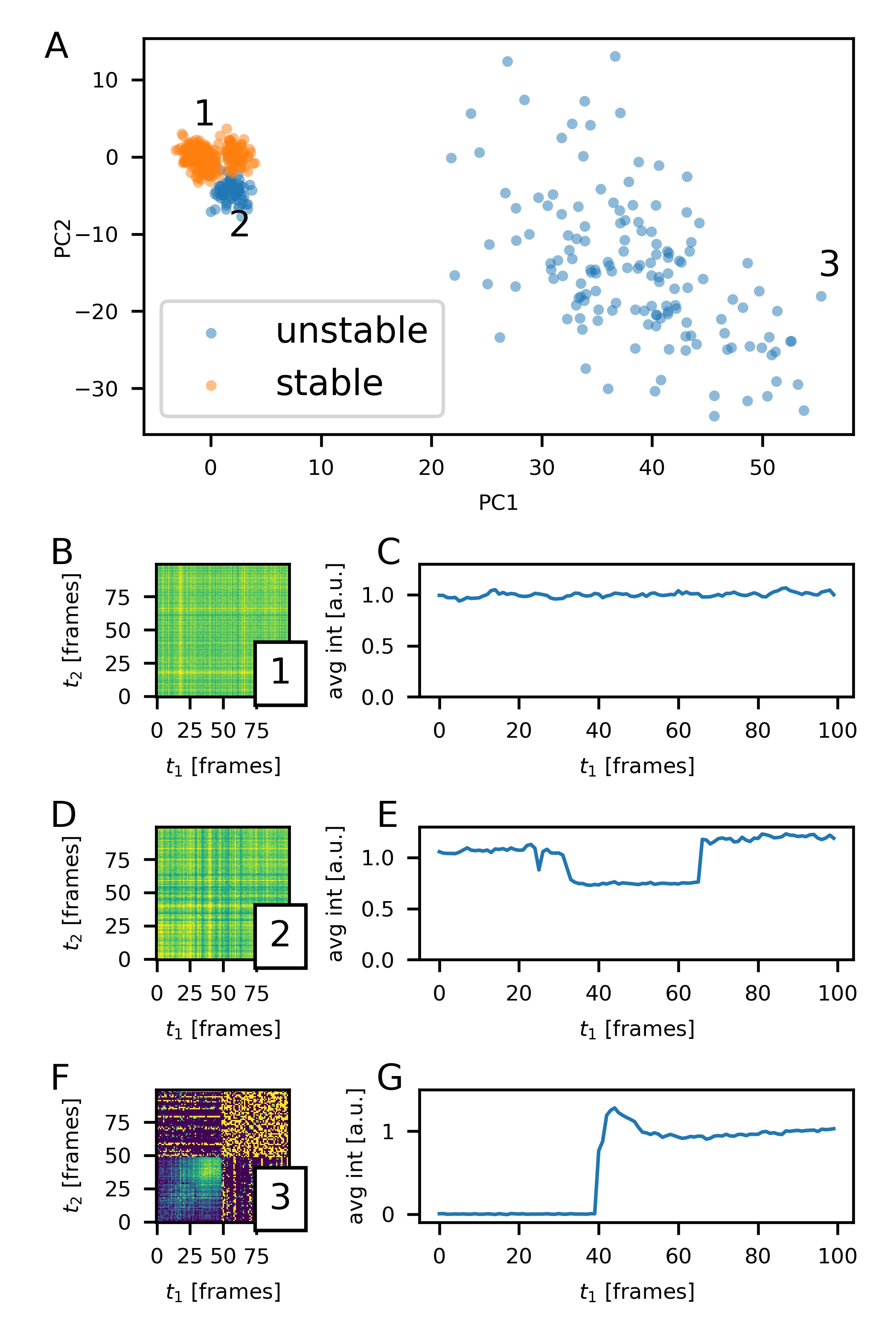}%
\caption{\label{fig:Figure6}Detection of anomalies in static experiments by encoding data with the denoising AE model. (A) points in the latent space corresponding to inputs generated from the experimental series. The points are mapped to 2D space using linear principal components. B,D,F – $2TCF$s for stable(B), slightly unstable(D) and very unstable(F) experimental conditions. C,E,F – average intensities on the detector for the respective experiments on the left.}
\end{figure*}

Figure~\ref{fig:Figure6} shows the representations of stable and unstable series in the latent space. Inputs formed from the same full-sized $2TCF$ are located in close proximity of each other, forming a tight cluster. This confirms the contractive property of the denoising AE - the fact that similar inputs are close in the latent space. Having multiple $2TCF$s representing the same dynamics through \emph{down-mapping} procedure allows estimating the characteristic size of the cluster, to which all other points in the latent space should be compared to. The means and variances of contrasts for the inputs collected from experiment with a small disturbance are similar to the undisturbed series, resulting in a small distance in the latent space, comparable to the size of the clusters. However, when the disturbance is so strong, that the beam is lost for part of the series, the latent space representations of a $2TCF$ are much further away from the undisturbed series. In fact, the strongly disturbed series form a separate cluster, where points are much more spread out than for stable series. Selecting a threshold distance between points helps to filter out the measurements taken at very unstable experimental conditions.  Thus, it is possible to identify unusual observations in a set of XPCS series by looking into the denoising AE’s latent representation of $2TCF$s and comparing the distance between series. 

\section{\label{sec:Discussion}Discussion}

Analysis of XCPS data is a multistep, often iterative, process that requires continuous evaluation of intermediate results by a domain expert. Large volumes of observations and subjectivity in alleviating low signal-to-noise ratio complicate proper extraction of valuable information from experiments. Computationally reducing noise in $2TCF$s as a data processing step helps to achieve several goals: automating the analysis workflow, improving temporal resolution of parameters that quantify the system’s dynamics and increasing quantitative usage of data with high cost of collection. 

Here, we demonstrate how a denoising AE model can be included into analysis workflow of XPCS data for experiments with ageing dynamics. Quantification of the model’s bias helps driving the decision about the use of its outputs as well as flagging unusual observations, such as heterogeneities or dynamics that are very different from ageing. The denoised correlation function can be used for complementing the fit of the raw data. The concept of \emph{trust regions} combines the assessment of the fit quality as well as domain expertise, which helps to not only report the most reliable results, but to automate sequential narrowing of the fit parameters boundaries.

The model's performance for unseen 3D printing data (not included in the training/validation datasets) demonstrates the advantages of its application to actual XPCS experiments. For measurements where the material’s dynamics are similar enough to the model’s training dataset, the analysis does not require a human-in-the-loop after all the thresholds are selected prior to analysis. Moreover, the analysis is suitable for autonomous data acquisition. It provides values of dynamics parameters, which are important for making decisions about adapting data acquisition parameters such as acquisition rate, exposure time, duration of data acquisition or about changing the sample and/or the processing parameters.  

We further demonstrate how encoded representations of a $2TCF$ can be used for quantitative comparison of two or more scattering time series, which can be useful for identifying anomalies such experimental instabilities or phase transitions. The comparison can even be done for the types of dynamics, not present in the training set.

In conclusion, in this work we demonstrate how a CNN-based denoising AE can be used for $2TCF$s with non-equilibrium dynamics for experimental scattering series of arbitrary size. Addition of the denoising AE to XPCS analysis along with quantifying uncertainty helps automating the analysis and improves temporal resolution of extracted parameters. Besides, analysis of residuals and latent space representations of the inputs helps detect anomalous dynamics that go beyond monotonic ageing. This property can be employed for recognizing heterogeneities or phase transitions. Several examples of incorporating the denoising AE in analysis workflows for actual experimental data, collected at NSLS-II, demonstrate its effectiveness for unseen data and diversity of its applications. Denoising and encoding properties of the model are auspicious for various online and offline data analysis tasks at XPCS facilities. 

\begin{acknowledgments}
This research used the CHX and CSX beamlines and resources of the National Synchrotron Light Source II, a U.S. Department of Energy (DOE) Office of Science User Facility operated for the DOE Office of Science by Brookhaven National Laboratory(BNL) under Contract No. DE-SC0012704 and under a BNL Laboratory Directed Research and Development (LDRD) project 20-038 "Machine Learning for Real-Time Data Fidelity, Healing, and Analysis for Coherent X-ray Synchrotron Data"
\end{acknowledgments}

\appendix

\section{\label{sec:Denoising Model Training}Denoising Model Training}
The denoising model is written in Pytorch (version 1.9.0+cuda10.2). The model is trained using CUDA accelerator Nvidia GeForce RTX 2070 Super. Early stopping is used for model overfitting. Aside a few changes, the same model architecture and training procedure is used as described in \cite{konstantinova_2021}. 

The model is trained in the autoencoder regime, i.e. each input serves as its target. The cost function is the mean square loss between the model’s output and its target(input). Adam \cite{Kingma_2014} optimizer is used for training. Its starting learning rate of 1e$^{-4}$ (batch size 8) is decreased at each epoch by a factor 0.9995.  

One training epoch takes from 28 seconds (kernel size 1) to 320 seconds (kernel size 17). 20-30 epochs are required for training, depending on random initialization of the model’s weights and the order of training examples supplied to the model.

\section{\label{sec:Fitting 2TCF}Fitting a $2TCF$}

Each individual $1TCF$ is fit to Eq.~\ref{eq:(3)} using nonlinear regression implemented in Scipy package \cite{2020SciPy-NMeth} according to the Trust Region Reflective algorithm \cite{ branch_1999}. Generally, when a $1TCF$ is calculated for a single-frame-wide cut of a $2TCF$, each of its points is given equal weight during the fit and when several-frames-wide cuts are considered, the weights are assigned inversely proportionally to the standard errors of correlation function values at each $t_d$ point. 

The optimization algorithm for nonlinear regression is sensitive to the initial guess of the parameters. The initial values for $\beta$ and $C_{\infty}$ can be estimated from characteristics of the experimental setup and test measurements. The algorithm seems to be robust with respect to the initial value of $\alpha$, which is set to 1. However, a proper initial guess of $\Gamma$ is important and  $\Gamma$ values  can vary significantly between experiments and ROIs in the reciprocal space. To automate selection of the initial guess for $\Gamma$, several values are attempted and the fit with the smallest $R^2$ is returned. The values range between 0.01 and 1 \emph{fr}. If neither of the values results in better fit than a line $1TCF(t_d) = const$, then the error (\emph{`not a number'}) is returned for all parameters. 

It is common to have a situation when a signal decorrelates within a small delay and most of the points along the $t_d$ axis in a $1TCF$ are at the baseline level. In such case, points at the tail of the $1TCF$ have high leverage, but carry little information regarding the dynamics' rate, the contrast factor and the compression constant. The equidistant delay points in $1TCF$s obtained through cuts of a $2TCF$ contrast with logarithmic-spaced delay points of $1TCF$s, calculated directly from the speckled images using the traditional multi-tau algorithm \cite{wohland_2001}. To reduce the influence of the points at the tail of the $1TCF$, lower weights can be assigned to them during the fit with nonlinear regression. For implementing this, at the first pass, the data are fit to Eq.~\ref{eq:(3)} with points weighted either equally or according to standard errors, as described above.
Based on the results of the first fit, the points for which the fitted values are larger than $1+\frac{\beta}{e}$ are assigned the weight of 1 during the second fit and the rest of the points are assigned the weight $\frac{1}{log(t_d)}$. The same data are fit for the second time to Eq.~\ref{eq:(3)} with the fixed value $C_{\infty}$ from the previous fit and the new points' weights. Both weighted and unweighted results are returned by the fitting procedure. 

The nonlinear regression algorithm returns the parameter values as well as their covariance matrix. The values of $\Gamma$ and $\alpha$ for the best fit are used for calculating the half-time (Eq.~\ref{eq:(4)}) and establishing initial \emph{trust regions}. Square roots of variances (the diagonal elements of the covariance matrix) are attributed to the parameters' errors. The off-diagonal elements are converted to correlations. Optionally, the parameters' errors and correlations and the $R^2$ values are used for narrowing the \emph{trust regions} based on user-defined thresholds.  

\section{\label{sec:Output Files Formats}Output Files Formats}

$2TCF_{denoised}$ and uncertainty scores are saved as datasets in a \emph{hdf5} \cite{hdf5_cite} file, which allows accessing data (loading into memory) in parts. Parameters, parameter’s errors, parameter’s correlations and the $R^2$ of the fits at each age are recorded as nested dictionaries into a \emph{json} \cite{json_cite} file. Besides, the parameter boundaries and the initial guesses are also saved in a \emph{json} file. 

\section{\label{sec:Different Kernel Sizes}Different Kernel Sizes}

\begin{figure*}
\includegraphics[width=\linewidth]{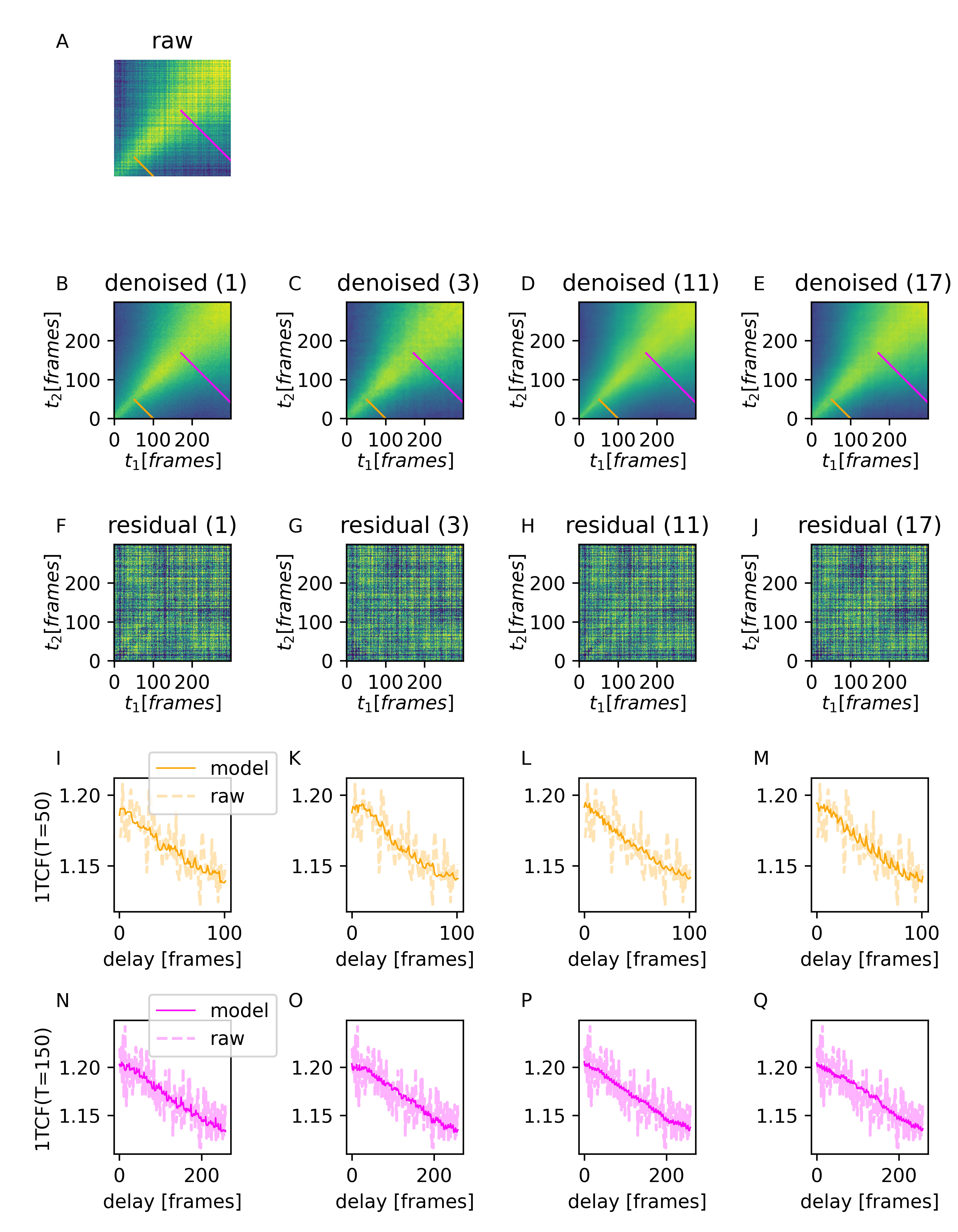}%
\caption{\label{fig:FigureS1} Comparison of denoising with models with different kernel sizes (1,3,11,17) for relatively slow dynamics. (A) $2TCF_{raw}$, (B-E) $2TCF_{denoised}$, (F-J) residuals, (I-M) $1TCF$s for cut at $t_a = 150 frames$, (N-Q) $1TCF$ for $t_a=50 frames$. $\Delta t_a = 1 frame$. Color scale is the same within each row.}
\end{figure*}

\begin{figure*}
\includegraphics[width=\linewidth]{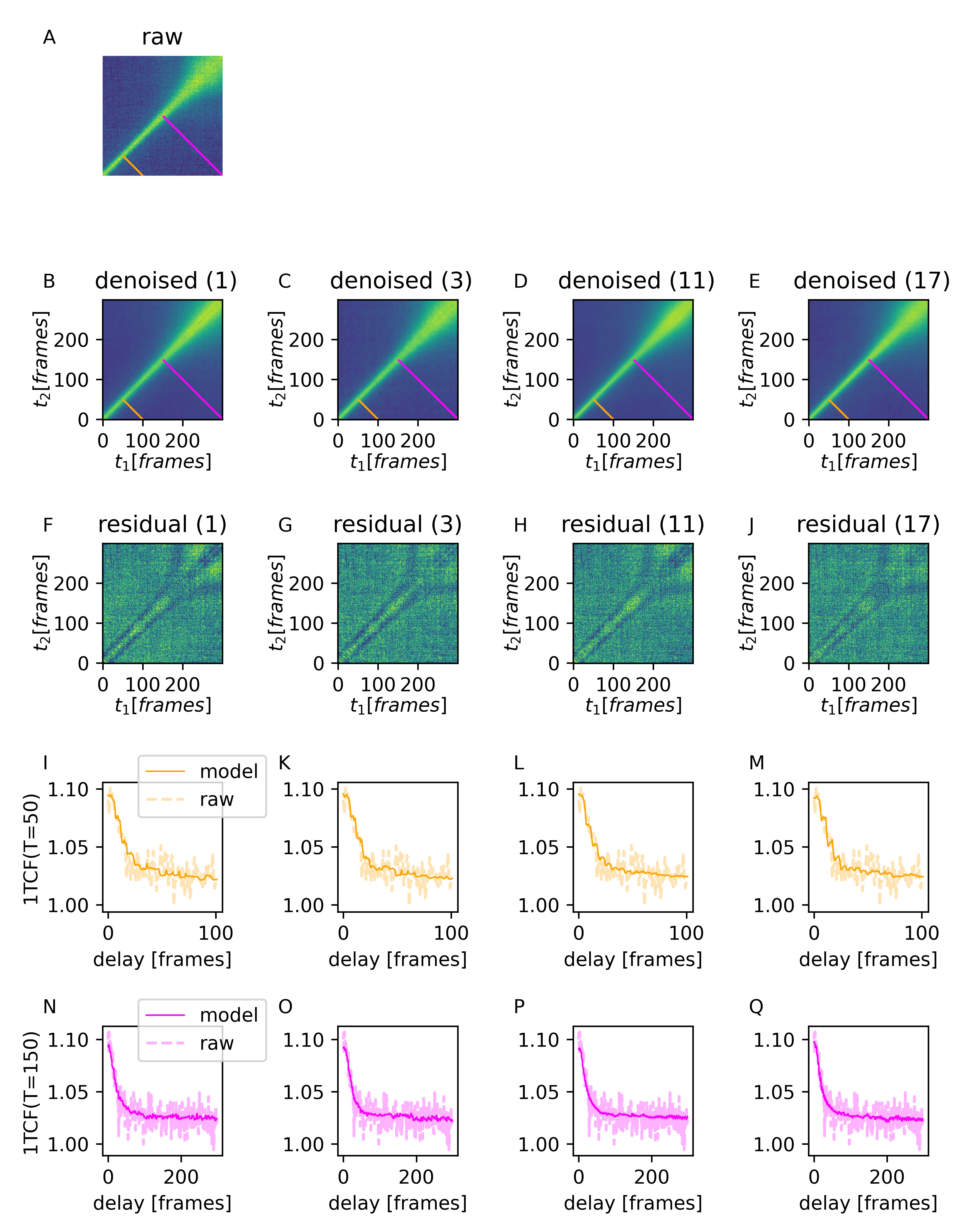}%
\caption{\label{fig:FigureS2} Comparison of denoising with models with different kernel sizes (1,3,11,17) for relatively fast dynamics. (A) $2TCF_{raw}$, (B-E) $2TCF_{denoised}$, (F-J) residuals, (I-M) $1TCF$s for cut at $t_a = 150 frames$, (N-Q) $1TCF$ for $t_a=50 frames$. $\Delta t_a = 1 frame$. Color scale is the same within each row.}
\end{figure*}

Comparison of the denoising AEs with different convolutional kernels shows that models with larger kernel sizes have less bias and closer resemble the dynamics even if they slightly deviate from the KWW form. Examples of model outputs for the validation data are shown in Fig.~\ref{fig:FigureS1} and Fig.~\ref{fig:FigureS2}. The figures show the model output for the $2TCF$s with 300 frames, rather than for single 100-frame inputs (see the main text for details of applying the model to an arbitrary-sized input). Generally, with increasing the kernel size of the model, the trends in the residuals become less pronounced. 

However, the difference between the models’ outputs is generally not significant for quantitative analysis, which involves further fitting of the $1TCF$s. We compare the error of parameters extracted from the $2TCF$s passed through the models with different kernels for all full-sized $2TCF$s from the validation set. For this, we fit single-frame-wide cuts of a $2TCF_{raw}$ and its denoised version $2TCF_{denoised}$ in an iterative process according to Fig.~\ref{fig:Figure4}. The initial fitting of $2TCF_{denoised}$ is used for establishing parameter boundaries for the contrast factor $\beta$ and the baseline $C_{\infty}$, applied during the second fit. The ground truth values are obtained by separately fitting 20-frames-wide cuts of $2TCF_{raw}$ and interpolating the results of the fit across the entire range of ages. The ground truth fits involve human intervention for establishing the parameters’ boundaries. 

One $2TCF$ is removed from consideration because its dynamics are too fast to be captured with the \emph{down-and-up mapping}  approach. Two other $2TCF$s are removed from consideration because the \emph{trust regions} for $\beta$ from the first fit of $2TCF_{denoised}$ contained less than two points and thus the parameter boundaries could not be established. Those $2TCF$s contained very slow dynamics.  Overall, 8100 single-frame $1TCF$s have been fit for each kernel value.

\begin{figure*}
\includegraphics[]{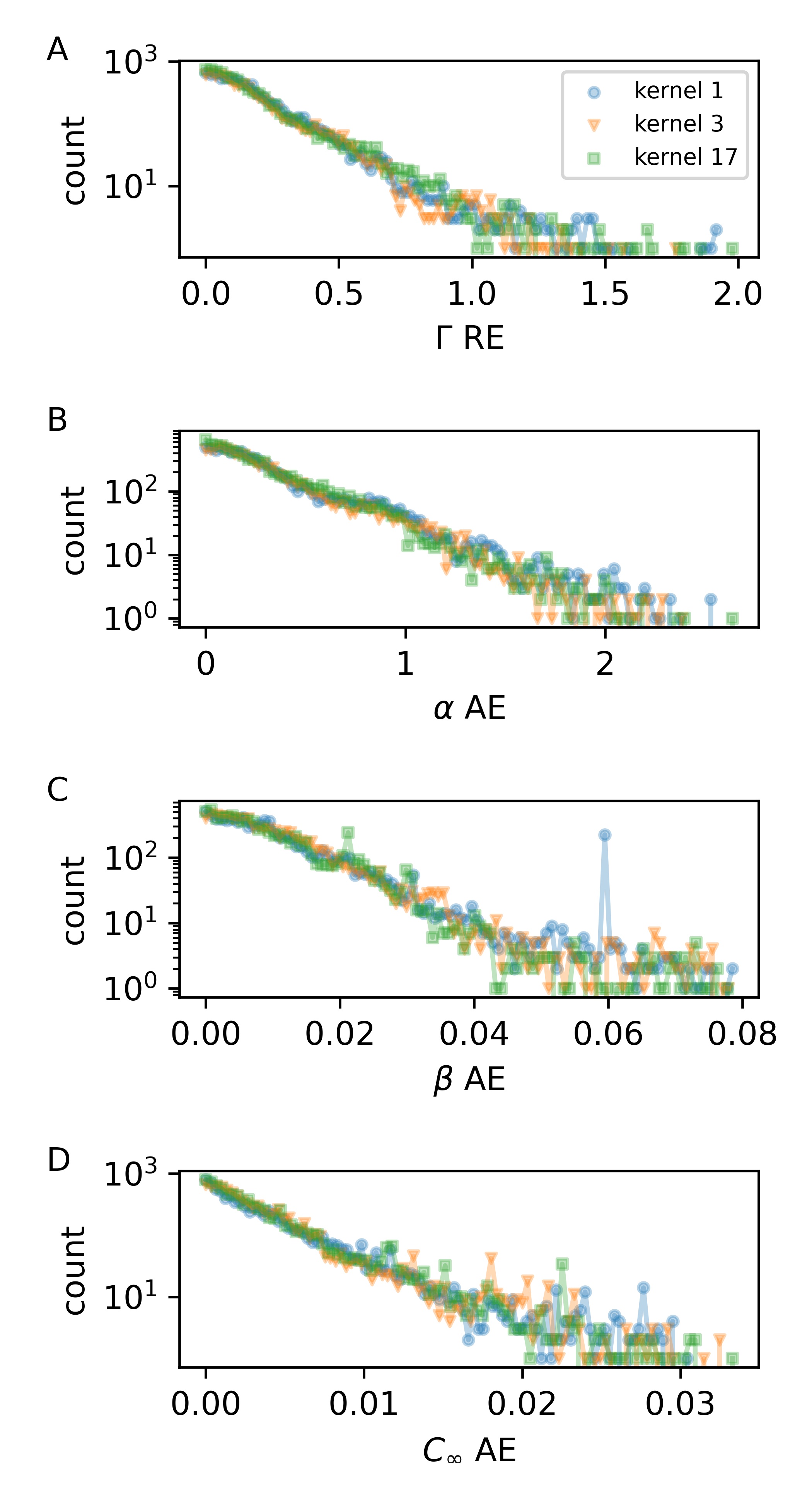}%
\caption{\label{fig:FigureS3} Distribution of errors for parameters extracted from $2TCF$s in the validation set, denoised with models having kernel size 1 (circle), 3(triangle) and 17(square). Relative error for dynamics rate $\Gamma$ (A); absolute error for stretching constant $\alpha$ (B), $\beta$ (C) and $C_{\infty}$ (D).}
\end{figure*}

The distribution of errors for the models with kernels 1, 3 and 17 is shown in Fig.~\ref{fig:FigureS3}. For $\Gamma$ the relative error (RE$\Gamma$) is considered since dynamics' rate can be in a broad range, including values close to zero. The absolute error is considered for $\alpha$ (AE$\alpha$), $\beta$ (AE$\beta$) and $C_{\infty}$ (AE$C_{\infty}$). Table~\ref{tab:errors} shows the comparison of median values of the errors. The analysis procedures involving all models produce comparable errors. 

\begin{table*}
\begin{tabularx}{\textwidth} { 
  | >{\raggedright\arraybackslash}X 
  | >{\centering\arraybackslash}X 
  | >{\centering\arraybackslash}X 
  | >{\centering\arraybackslash}X
  | >{\centering\arraybackslash}X 
  | >{\raggedleft\arraybackslash}X | }
 \hline
 kernel & RE$\Gamma$(model) & AE$\beta$(model) & AE$C_{\infty}$(model)& AE$\alpha$(model)\\
 \hline
  1 & 0.133 & 0.0076 & 0.0023 & 0.21\\
 \hline
  3 & 0.135 & 0.0081 & 0.0026 & 0.23\\
 \hline
  7 & 0.127 & 0.0075 & 0.0023 & 0.21\\
   \hline
  11 & 0.121 & 0.0069 & 0.0023 & 0.21\\
   \hline
  17 & 0.132 & 0.0075 & 0.0025 & 0.22\\
 \hline
 \hline
 kernel & RE$\Gamma$(raw) & AE$\beta$(raw) & AE$C_{\infty}$(raw)& AE$\alpha$(raw)\\
 \hline
  NA & 0.195 & 0.0088 & 0.0025 & 0.35\\
\hline

\end{tabularx}

\caption{\label{tab:errors} Errors for parameters extracted from a $2TCF$ denoised with different kernels and from the raw data for examples from the validation set.}
\end{table*}

The errors produced by fitting $2TCF_{raw}$ have larger median errors for both RE$\Gamma$ and AE$\alpha$, even when the parameter bounds for $\beta$ and $C_{\infty}$ are narrowed. The difference in errors for $\beta$ and $C_{\infty}$ between fits for $2TCF_{raw}$ and $2TCF_{denoised}$ is not significant because in both cases the parameter boundaries were established using $2TCF_{denoised}$ at the previous step.  

\begin{figure*}
\includegraphics[width=\linewidth]{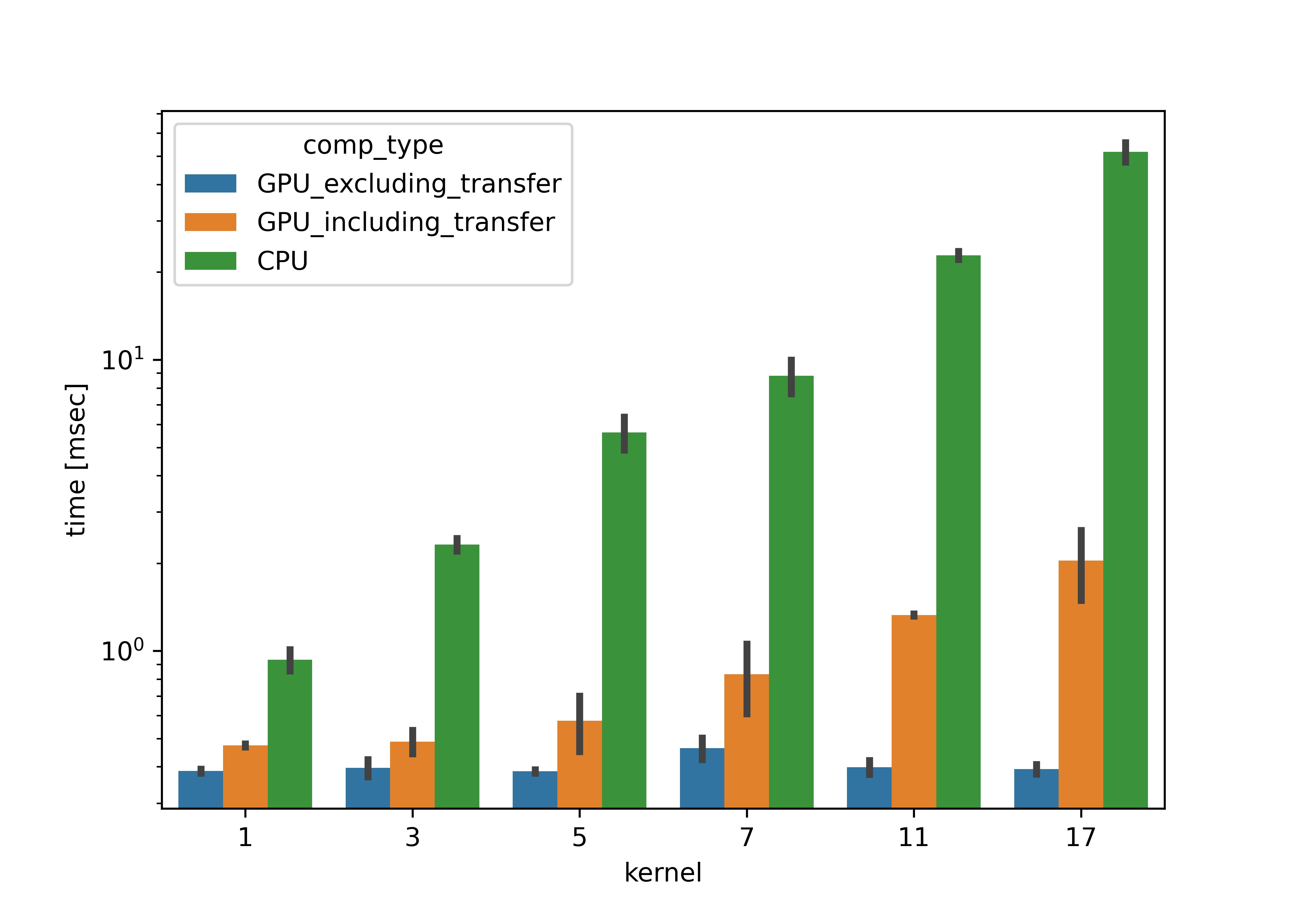}%
\caption{\label{fig:FigureS4} Computation times for application of the denoising AE model (single 100x100 frames input) with/without GPU acceleration and with/without data transfer from CPU-based numpy arrays to CUDA-based Pytorch tensors. Error bars are standard deviations for measurements repeated 50 times.}
\end{figure*}

When considering denoising AEs with different kernel sizes, one needs to account for the computational time. The comparison of computational times for denoising models with different kernel sizes is shown in Fig.~\ref{fig:FigureS4}. These times are measured for a single 100-frame $2TCF$ input (average of 50 repeated measurements). Depending on the acceleration hardware and the need to transfer data to/from a CPU/GPU Pytorch Tensor, the difference between computation times for models with kernel=1 and kernel=17 can be 60 times (CPU) or can be non-existent (GPU, without transfer).  

\section{\label{sec:Model Variance}Model Variance}

We estimate the variance of a denoising AE model by applying an ensemble of 9 models trained with different random initializations to examples from the validation set.  For each full-sized $2TCF$ from the validation set, a reduced version $2TCF_{reduced}$ is generated by considering each 3rd frame horizontally and vertically, starting at point [0,0], and then cropping the first 100x100 frames part. The resulting $2TCF_{reduced}$ is passed to each of the models in the ensemble and the frame-wise variance ($\sigma^2$) is calculated. That is, each of the 100$\times$100 frames input produces $10^4$ variances. We refer to these variances as generated via drawing from \emph{model space}.  

When a single denoising AE model is applied to a $2TCF$ with 200 or more frames, the \emph{down-and-up mapping} process also introduces variance between neighboring values. The difference in underlying dynamics represented in the \emph{down-mapped} $2TCF$ originating from starting points [i, i] (i, j = 0,1,2) is often negligible. Therefore, the \emph{down-mapped} $2TCF$s can be viewed as an approximation of realization of the same dynamics. Thus, it is possible to estimate the variance of a model via drawing from \emph{input space} by applying the model to each of such realizations and estimating the variances of the outputs. Here, we estimate the variance by considering inputs, obtained from the $2TCF$s from the validation set by considering each 3\textsuperscript{rd} frame horizontally and vertically, starting at points: [0,0], [1,1], [2,2] and then cropping the first 100$\times$100 frames part from each of the \emph{down-mapped} $2TCF$. A single model is applied to each of the 3 inputs and the frame-wise variance is calculated.

\begin{figure*}
\includegraphics[width=\linewidth]{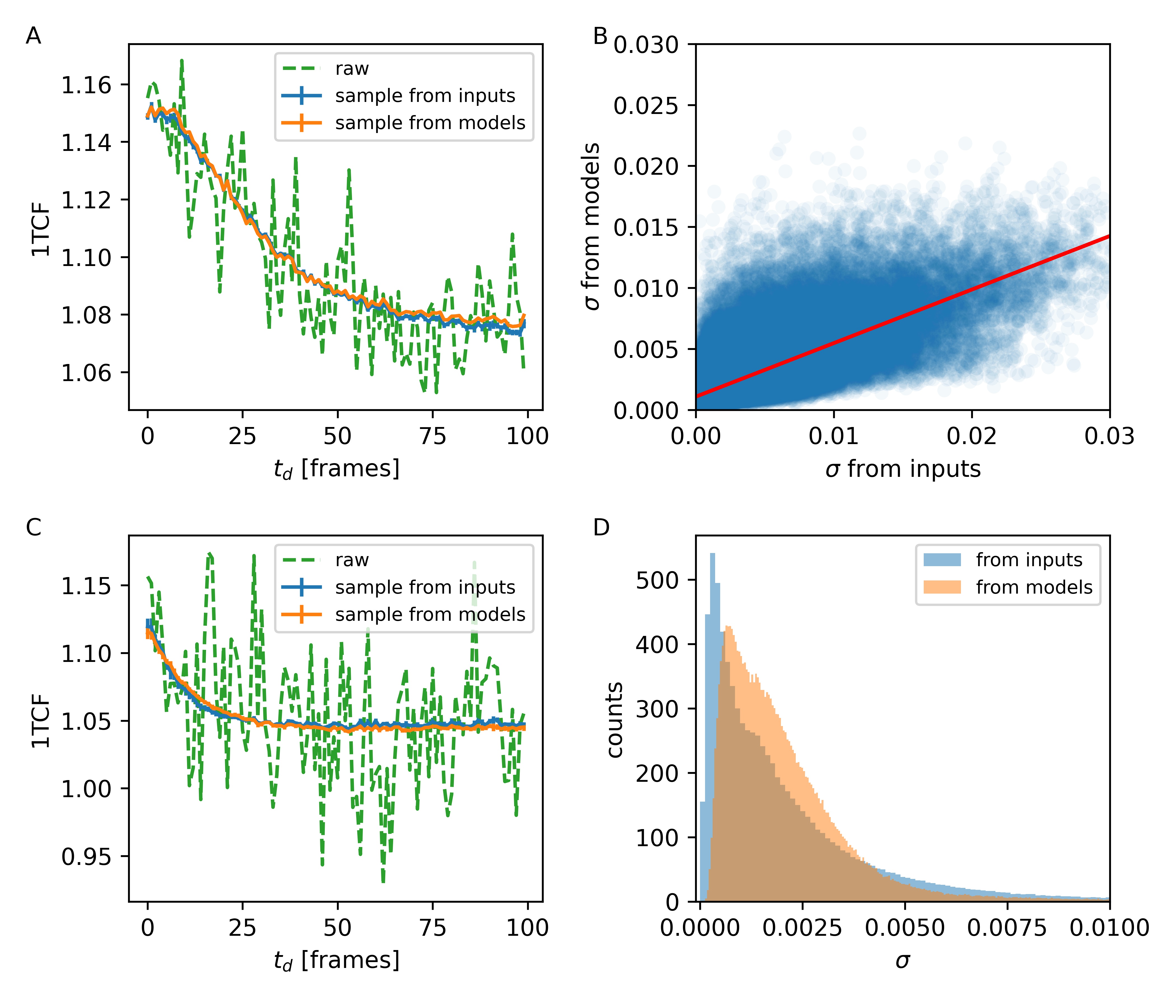}%
\caption{\label{fig:FigureS5} Variance of the model. (A,C) A $1TCF$ with $\Delta t_a=1 fr^{-1}$ obtained from a $2TCF_{raw}$ (green) of two different experiments and from respective average of $2TCF_{denoised}$s for ensembles drawn from the \emph{model space} (orange) and the \emph{input space} (blue). (B) Standard deviations of $2TCF_{denoised}$ values for ensembles drawn from \emph{model space} versus respective values for ensembles drawn from the \emph{input space}. Each point corresponds to a value of a $2TCF$ at a single point $[t_1, t_2]$. Red line is a linear fit between the variables. (D) Distribution of the standard deviation values from (B).}
\end{figure*}

As a result, for each example in the validation set the frame-wise variance is calculated by  drawing from both \emph{model space} and \emph{input space}. The variances, calculated by both methods, are comparable. (Fig.~\ref{fig:FigureS5}) The variance values are considerably smaller than both the noise in the original raw inputs and the typical values of contrast factor $\beta$, making the model's variance negligible for the quantitative analysis. 

Moreover, there is a linear dependence between the variances calculated by the two methods (the correlation coefficient is 0.73), indicating that the model variance typically is already reflected in the \emph{down-and-up mapping} process of the model application. Therefore, it is not necessary to separately estimate the variance of the model using an ensemble method. 

\section{\label{sec:Latent Coordinates Distribution}Latent Coordinates Distribution }

\begin{figure*}
\includegraphics[width=\linewidth]{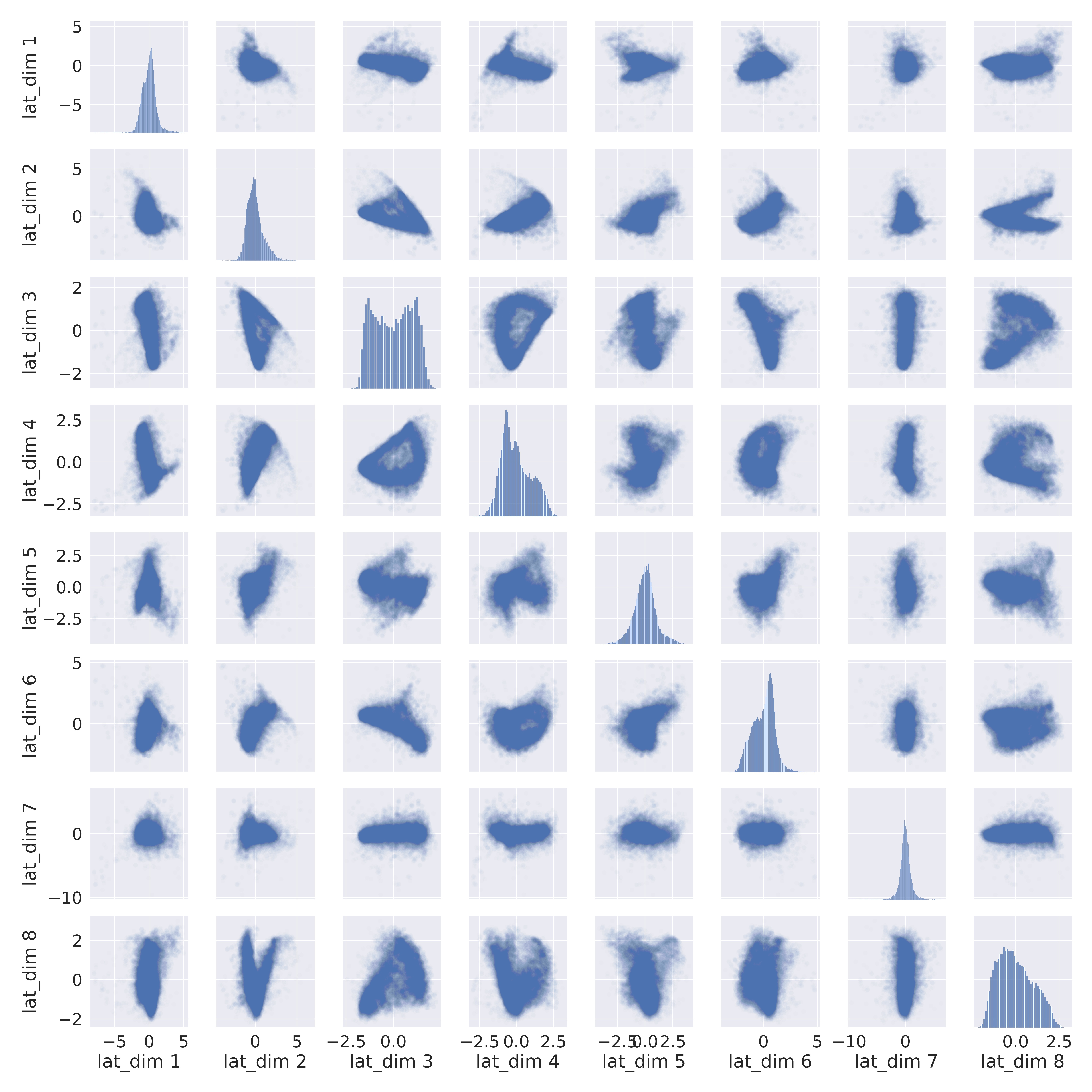}%
\caption{\label{fig:FigureS6}Pairwise distribution of latent coordinates for the training set. The coordinates are scaled to have zero mean and unit variance.}
\end{figure*}

Latent space coordinates can be used for applying distance-based similarity measures for the model inputs. The model is likely to perform well for new inputs that are similar to the training set examples. The pair-wise distributions of the latent coordinates for the training set (Fig.~\ref{fig:FigureS6}) reveals that the majority of the data is approximately centered around a single point. As a result, the measure of similarity between a new input and the training set can be expressed as the Euclidean distance from the input to the central point of the training set coordinates distribution. Prior to calculation of the distance, the coordinates are standardized using the mean values and variances for each coordinate calculated for the training set. 

\section{\label{sec:Quantifying Bias}Quantifying Bias With Residuals}

For general applications, investigating trends in residuals is a common test for a model’s bias. If a residual resembles a ‘random’ noise, the bias of the model is low. In an opposite case of trends in the residual, the model does not fully represent the process in the input, i.e., it exhibits a bias. 

\begin{figure*}
\includegraphics[width=\linewidth]{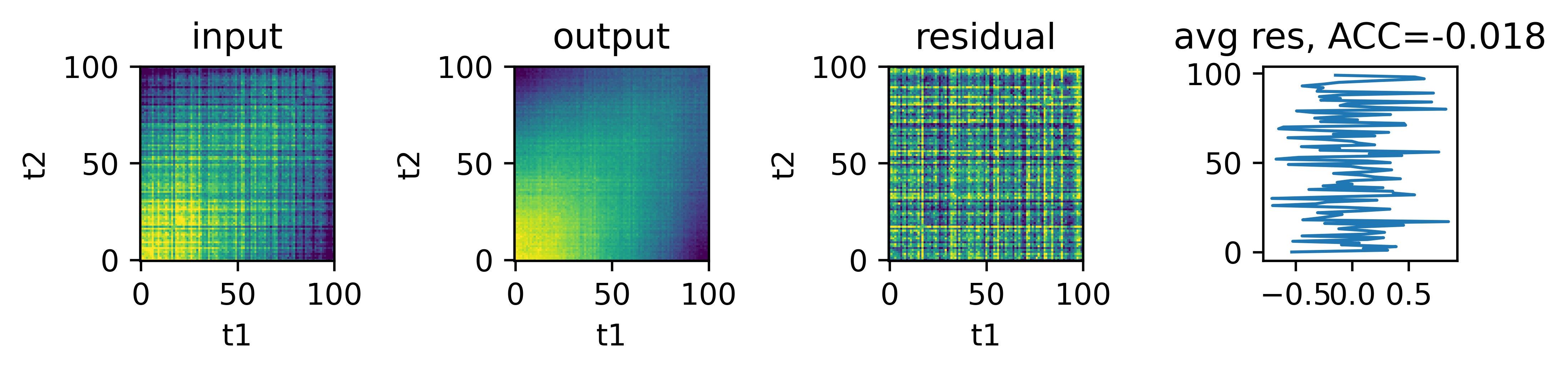}%
\caption{\label{fig:FigureS7}Example from training set where the residual does not have trends.}
\end{figure*}

\begin{figure*}
\includegraphics[width=\linewidth]{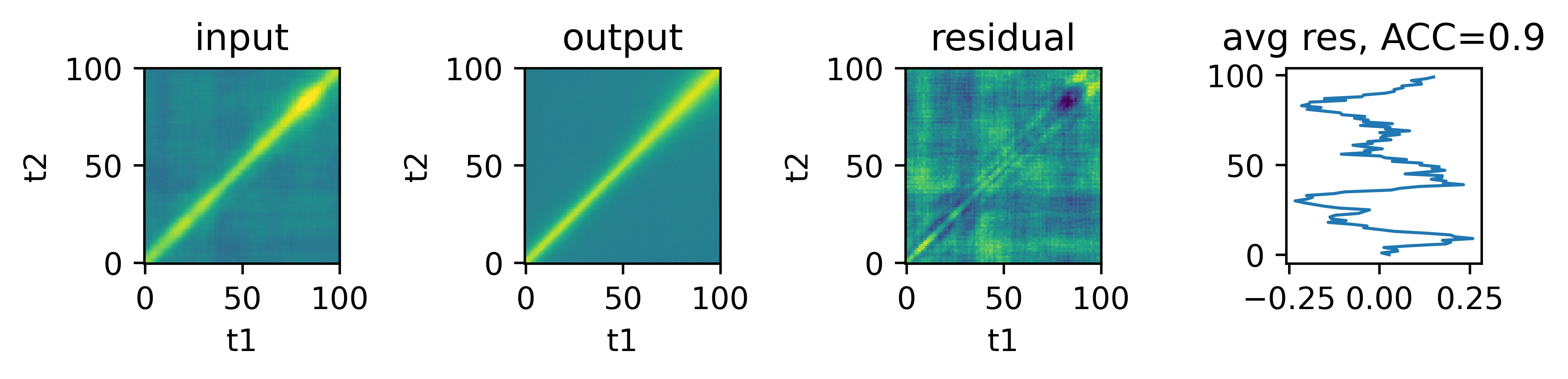}%
\caption{\label{fig:FigureS8}Example from training set where the residual has a well pronounced trend.}
\end{figure*}

For a denoising AE the situation is not straightforward and depends on the perception of noise for each application. For a $2TCF$, in cases of low detector count or frame-by-frame instabilities, the residual for the denoised output should not have any trends (Fig.~\ref{fig:FigureS7}). However, in cases when an average, ‘envelope’, dynamics is under interest, the dynamics heterogeneities are treated as noise, leaving a pattern in the residual (Fig.~\ref{fig:FigureS8}), that can be investigated separately. It is convenient to measure noise using autocorrelation coefficient (ACC) at lag=1  of residuals, projected at one of the time axes . 

\begin{figure*}
\includegraphics[width=\linewidth]{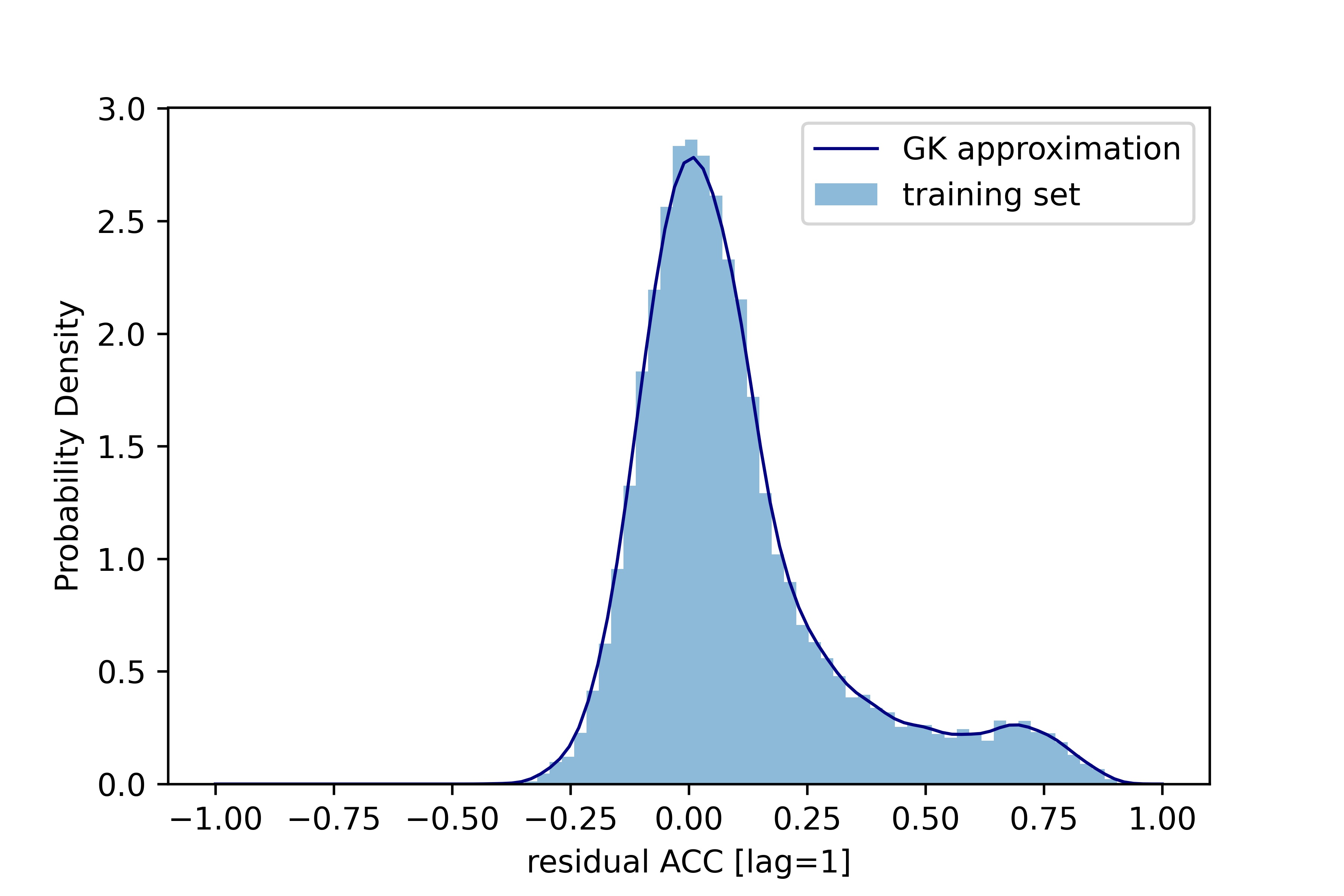}%
\caption{\label{fig:FigureS9}Distribution of the model uncertainty (ACC(lag=1) of residuals) for the training set. Solid blue line is the probability density function for the distribution calculated with Gaussian Kernel density estimator.}
\end{figure*}

Just like the latent space coordinates, the ACC of residuals for each new example can be compared to the analogous values for examples in the training set. This helps to understand, what values of ACC are unusual. For quantifying the comparison, the probability density function (Fig.~\ref{fig:FigureS9}) is estimated for ACC values of the training set and is calculated for each new test example.  

\section{\label{sec:Algorithms}Examples of workflows for Online and Offline Data Analysis}

Algorithm 1 (online analysis): 

\begin{itemize}
  \item apply denoising AE to the $2TCF_{raw}$ to obtain $2TCF_{denoised}$ 
  \item calculate an uncertainty score through latent space and/or residuals
  \item unless observation is unusual:
    \begin{itemize}
        \item do the first round of fit for $2TCF_{denoised}$ (optionally with wider time cuts) using wide parameter boundaries
        \item calculate new parameter boundaries based on values within \emph{trust regions}
        \item if needed, repeat the fits for $2TCF_{denoised}$ for single-frames cuts with new parameter boundaries 
    \end{itemize}
    \item if the observation is unusual:
    \begin{itemize}
        \item use $2TCF_{raw}$ and wide-bin $1TCF$s for the fits
    \end{itemize}
    \item report results of the fit and \emph{trust regions}
\end{itemize}

Algorithm 1 (offline analysis): 
\begin{itemize}
  \item apply denoising AE to the $2TCF_{raw}$ to obtain $2TCF_{denoised}$ 
  \item calculate an uncertainty score through latent space and/or residuals
  \item based on the uncertainty score, alarm about an unusual observation if encountered 
  \item unless observation is unusual:
    \begin{itemize}
        \item do the first round of fit for $2TCF_{denoised}$ (optionally with wider time cuts) using wide parameter boundaries
        \item calculate new parameter boundaries based on values within \emph{trust regions}
        \item fit both $2TCF_{raw}$ and $2TCF_{denoised}$ with new parameter bounds 
    \end{itemize}
    \item if the observation is unusual:
    \begin{itemize}
        \item use $2TCF_{raw}$ and wide-bin $1TCF$s for the fits
        \item report results of the fit and \emph{trust regions}
    \end{itemize}
\end{itemize}

\clearpage
%
\end{document}